\documentclass[prl,twocolumn,showpacs,amsmath,amssymb]{revtex4}

\usepackage{graphicx}
\usepackage{dcolumn}
\usepackage{bm}

\begin{document}

\title{Counting statistics of single-electron transport in a quantum dot}

\author{S. Gustavsson} \email{simongus@phys.ethz.ch}
 \affiliation{Solid State Physics Laboratory, ETH Z\"urich, CH-8093 Z\"urich, Switzerland}

\author{R. Leturcq}
\affiliation{Solid State Physics Laboratory, ETH Z\"urich, CH-8093 Z\"urich, Switzerland}

\author{B. Simovi\v c}
\affiliation{Solid State Physics Laboratory, ETH Z\"urich, CH-8093 Z\"urich, Switzerland}

\author{R. Schleser}
\affiliation{Solid State Physics Laboratory, ETH Z\"urich, CH-8093 Z\"urich, Switzerland}

\author{T. Ihn}
\affiliation{Solid State Physics Laboratory, ETH Z\"urich, CH-8093 Z\"urich, Switzerland}

\author{P. Studerus}
\affiliation{Solid State Physics Laboratory, ETH Z\"urich, CH-8093 Z\"urich, Switzerland}

\author{K. Ensslin}
\affiliation{Solid State Physics Laboratory, ETH Z\"urich, CH-8093 Z\"urich, Switzerland}

\author{D. C. Driscoll}
\affiliation{Materials Department, University of California, Santa
Barbara, CA-93106, USA}

\author{A. C. Gossard}
\affiliation{Materials Department, University of California, Santa
Barbara, CA-93106, USA}

\date{\today}

\begin{abstract}
We have measured the full counting statistics (FCS) of current
fluctuations in a semiconductor quantum dot (QD) by real-time
detection of single electron tunneling with a quantum point contact
(QPC). This method gives direct access to the distribution function
of current fluctuations. Suppression of the second moment (related
to the shot noise) and the third moment (related to the asymmetry of
the distribution) in a tunable semiconductor QD is demonstrated
experimentally. With this method we demonstrate the ability to
measure very low current and noise levels.
\end{abstract}

\maketitle

Current fluctuations in conductors have been extensively studied
because they provide additional information compared to the average
current, in particular for interacting systems \cite{Blanter02}.
Shot noise measurements demonstrated the charge of quasiparticles in
the fractional quantum Hall effect \cite{PicciottoSaminadayar} and
in superconductors \cite{Jehl01}. However, to perform such
measurements for semiconductor quantum dots (QD) using conventional
noise measurements techniques is very challenging. This is because
of the very low currents and the corresponding low noise levels in
these systems. Earlier experiments demonstrated the measurement of
shot noise in non-tunable QDs \cite{Birk02,Nauenetal}, but to our
knowledge, no experiments have been reported in the literature in
which the tunnel barriers, and thereby the coupling symmetry, could
be controlled \cite{Kouw01}.

An alternative way to investigate current fluctuations, introduced
by Levitov {\it et al.}, is known as full counting statistics (FCS)
\cite{Levitov01}. This method relies on the evaluation of the
probability distribution function of the number of electrons
transferred through a conductor within a given time period. In
addition to the current and the shot noise, which are the first and
second moments of this distribution, this method gives access to
higher order moments. Of particular interest is the third moment
(skewness), which is due to breaking the time reversal symmetry at
finite current. Experimentally, few attempts to measure the third
moment have been made in tunnel junctions \cite{ReuletBomze}.

The most intuitive method for measuring the FCS of electron
transport is to count electrons passing one by one through the
conductor. Real-time detection of single electron transport has been
experimentally investigated only very recently
\cite{LuW01,Fujisawa01,Bylander01}. It is a challenging task since
it requires a very sensitive, low-noise and non-invasive
electrometer, as well as a high-bandwidth circuit. Several devices,
such as the single electron transistor \cite{LuW01,Fujisawa01} and
the quantum point contact (QPC)
\cite{Field01,Sprinzak01,Elzerman03,Schleser01,Vandersypen02}, have
been demonstrated to have high enough sensitivity to detect single
electrons in a QD. But, up to now, none of these experiments were
able to extract the full counting statistics of electron transport.

Here we report on the real-time detection of single electron
tunneling through a QD using a QPC as a charge detector. With this
method, we can directly measure the distribution function of current
fluctuations in the QD by counting electrons. To our knowledge, this
is the first measurement of the full counting statistics for
electrons in a solid state device. In addition, we can tune the
coupling of the QD with both leads and measure the respective
tunneling rates. We show experimentally the suppression of the
second and third moments of the current fluctuations when the QD is
symmetrically coupled to the leads.

\begin{figure}
\includegraphics[width=1\linewidth]{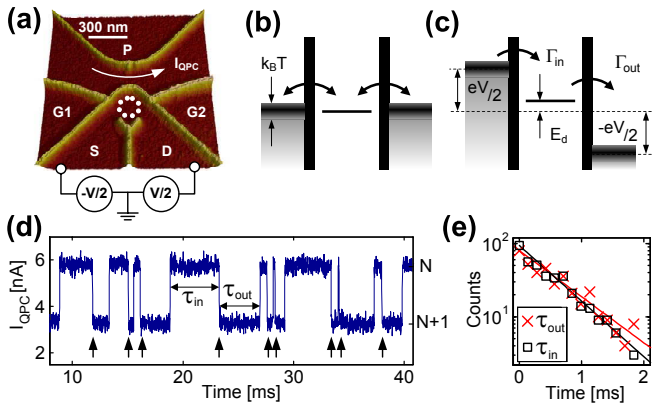}
\caption{\label{fig1} (a) AFM micrograph of the sample consisting of
a QD connected to two contacts $S$ and $D$, and a nearby QPC. G1, G2
and P are lateral gates allowing the tuning of the tunnel coupling
to the source $S$, the coupling to the drain $D$, and the
conductance of the QPC. G1 and G2 are also used to tune the number
of electrons in the QD. A symmetric bias voltage $V$ is applied
between the source and the drain on the QD. (b-c) Scheme of the
quantum dot in the case of equilibrium charge fluctuations (b), and
non-equilibrium charge fluctuations (c). (d) Time trace of the
current measured through the QPC corresponding to fluctuations of
the charge of the dot between $N$ and $N+1$ electrons. The arrows
indicate transitions where an electron is entering the QD from the
source lead. (e) Probability density of the times $\tau_{in}$ and
$\tau_{out}$ (see text) obtained from time traces similar to the one
in (d). The lines correspond to the expected exponential dependence
(see the text), where the tunneling rates are calculated from
$1/\Gamma_{S(D)} = 1/\Gamma_{in(out)} = \langle \tau_{in(out)}
\rangle$.}
\end{figure}

Figure~\ref{fig1}(a) shows the structure, fabricated on a
GaAs-GaAlAs heterostructure containing a two-dimensional electron
gas 34 nm below the surface (density $4.5 \times 10^{15}$ m$^{-2}$,
mobility 25 m$^2$(Vs)$^{-1}$). An atomic force microscope (AFM) was
used to oxidize locally the surface, thereby defining depleted
regions below the oxide lines \cite{Held02,Fuhrer04}. The
measurements were performed in a $^3$He/$^4$He dilution refrigerator
with an electron temperature of about 350 mK, as determined from the
width of thermally broadened Coulomb blockade resonances
\cite{Kouw01}. The charging energy of the QD is $2.1$ meV and the
mean level spacing is $200-300$ $\mu$eV. The conductance of the QPC,
$G_{QPC}$, was tuned close to $0.25 \times e^2/h$. We apply a dc
bias voltage between source and drain of the QPC, $V_{QPC}=500$
$\mu$V, and measure the current through the QPC, $I_{QPC}$, which
depends on the number of electrons $N$ in the QD.

In order to measure the current with a charge detector, one has to
avoid that electrons travel back and forth between the dot and one
lead or to the other lead due to thermal fluctuations
[Fig.~\ref{fig1}(b)]. This is achieved by applying a large bias
voltage between source and drain, i.e. $|\!\: \pm eV/2 - E_d| \gg
k_B T$, where $E_d$ is the electrochemical potential of the dot and
$V$ is the symmetrically applied bias, see Fig. \ref{fig1}(a, c). An
example of a time trace of the QPC current in this configuration is
shown in Fig.~\ref{fig1}(d). The number of electrons in the QD
fluctuates between $N$ and $N+1$. Since this trace corresponds to
the non-equilibrium regime, we can attribute each transition $N
\rightarrow N+1$ to an electron entering the QD from the source
contact, and each transition $N+1 \rightarrow N$ to an electron
leaving the QD to the drain contact. The charge fluctuations in the
QD correspond to a non-equilibrium process, and are directly related
to the current through the dot \cite{Fujisawa01}. Due to Coulomb
blockade, only one electron at a time can enter the QD, which allows
to count electrons traveling through the system.

\begin{figure}
\includegraphics[width=1\linewidth]{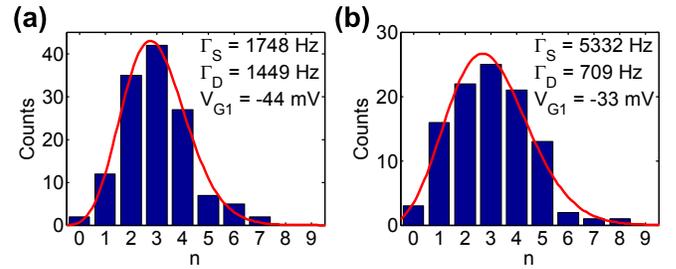}
\caption{\label{fig2} Statistical distribution of the number $n$ of
electrons entering the QD during a given time $t_0$. The two panels
correspond to two different values of the tunneling rates, obtained
for different values of the gate voltage $V_{G1}$. The time $t_0$ is
chosen in order to have the same mean value of number of events,
$\langle n \rangle \approx 3$, for both graphs. We have checked that
this choice does not affect the results. The line shows the
theoretical distribution calculated from Eqs.~(\ref{distribution})
and (\ref{generating}). The tunneling rates are determined
experimentally by the method described in Fig.~\ref{fig1}(e), and no
fitting parameters have been used for the theoretical curves.}
\end{figure}


The first application of electron counting in the non-equilibrium
regime concerns the determination of the individual tunneling rates
from the source to the QD, $\Gamma_S$, and from the QD to the drain,
$\Gamma_D$. Previous experiments determining the individual
tunneling rates involved more than two leads connected to the QD
\cite{Leturcq04}. In the trace of Fig.~\ref{fig1}(d), the time
$\tau_{in}$ corresponds to the time it takes for an electron to
enter the QD from the source contact, and $\tau_{out}$ to the time
it takes for the electron to leave the QD to the drain contact. For
independent tunneling events, the tunneling rates can be calculated
from the average of $\tau_{in}$ and $\tau_{out}$ on a long time
trace \cite{Schleser01}, $1/\Gamma_{S(D)} = 1/\Gamma_{in(out)} =
\langle \tau_{in(out)} \rangle$. To check that the tunneling events
are indeed independent, we have compared the probability densities
$p_{\tau_{in}}$ and $p_{\tau_{out}}$ with the expected exponential
behavior $p(\tau_{in(out)}) = \Gamma_{S(D)} \exp (-\Gamma_{S(D)}
\tau_{in(out)})$. Figure~\ref{fig1}(e) shows good agreement with our
data. It is interesting to note that, in the case shown in
Fig.~\ref{fig1}(e), the QD is almost symmetrically coupled. We
demonstrate here a very sensitive method to determine the symmetry
of the coupling alternative to Ref.~\onlinecite{Rogge02}.

\begin{figure}
\includegraphics[width=1\linewidth]{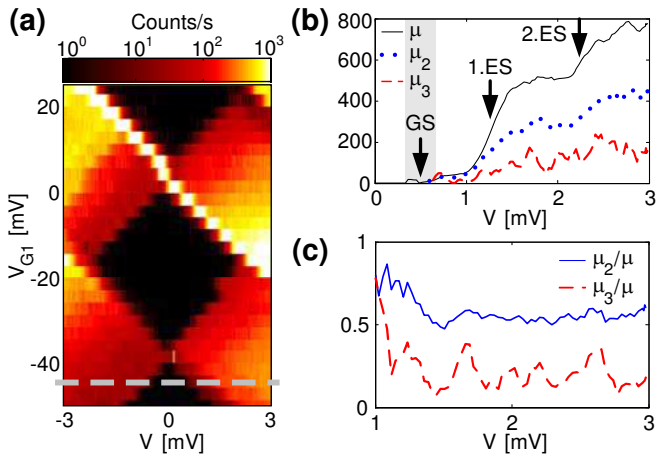}
\caption{\label{fig3} (a) Average number of electrons entering the
QD, $\mu$, measured as a function of the gate voltage $V_{G1}$ and
the bias voltage $V$. Far from the edges of the Coulomb blockade
region, i.e. for $|\pm eV/2 - E_d| \gg k_B T$, the fluctuations of
$n$ are directly related to current fluctuations. The dashed line
correspond to the cross-section shown in panel (b). (b) Three first
moments of the fluctuations of $n$ as a function of the bias voltage
$V$ and at a given gate voltage $V_{G1}=-44$ mV. The ground state
(GS) as well as two excited states (ES) are clearly visible. The
moments are scaled so that $\mu$ corresponds to the number of
electrons entering the QD per second. In the gray region, the
condition $|\pm eV/2 - E_d| \gg k_B T$ is not valid, and the number
of electrons entering the QD cannot be taken as the current flowing
through the QD. The width of this region is $9 \times k_BT/e \approx
300$ $\mu$V, determined from the width for which the Fermi
distribution is between 0.01 and 0.99. (c) Normalized second and
third moments as a function of the bias voltage $V$ and at a given
gate voltage $V_{G1}=-44$ mV.}
\end{figure}


From traces similar to the one shown in Fig.~\ref{fig1}(d), we can
directly determine the statistical properties of sequential electron
transport through the QD. We count the number $n$ of electrons
entering the QD from the source contact during a time period $t_0$,
i.e. the number of down-steps in Fig.~\ref{fig1}(d) (see arrows). We
obtain the distribution function of $n$ by repeating this counting
procedure on $m = T/t_0$ independent traces, $T=0.5$ s being the
total length of the time trace. The resulting distribution functions
are shown for two different values of the tunneling rates in
Figs.~\ref{fig2}(a) and \ref{fig2}(b).

The FCS theory allows to determine the distribution function of the
number $n$ of electrons traveling through a conductor
\cite{Levitov01}:
\begin{equation}
P(n) = \int_{-\pi}^{\pi} \frac{d\chi}{2\pi} e^{-S(\chi)-n\chi}, \label{distribution}
\end{equation}
where $S(\chi)$ is the generating function, which has been
calculated for a single level QD for large bias voltage $|\pm eV/2 -
E_d| \gg k_B T$ \cite{Bagrets01}:
\begin{eqnarray}
\frac{S(\chi)}{t_0} = \left[ \Gamma_S + \Gamma_D - \sqrt{\left( \Gamma_S - \Gamma_D \right)^2 + 4 \Gamma_S \Gamma_D e^{-i\chi}} \right] \label{generating}
\end{eqnarray}
Here $\Gamma_S$ and $\Gamma_D$ are the effective tunneling rates,
which take into account any possible spin degeneracy of the levels
in the QD, and correspond to the tunneling rates we determine
experimentally. We have calculated the theoretical distribution
functions for the tunneling rates measured in the cases of
Figs.~\ref{fig2}(a) and \ref{fig2}(b) [solid lines]. The agreement
with the experimental distribution is very good, in particular,
given that no fitting parameters were used. Both graphs show a clear
qualitative difference: Figure~\ref{fig2}(b) shows a broader and
more asymmetric distribution than Fig.~\ref{fig2}(a). We will see
later that this difference comes from the different asymmetries of
the tunneling rates.


In order to perform a more quantitative analysis, we calculate the
three first central moments given by $\mu = \langle n \rangle$, and
$\mu_i= \langle (n- \langle n \rangle)^i \rangle$ for $i=$2,3, where
$\langle ... \rangle$ represents the mean over $T/t_0$ periods of
length $t_0$. The first moment (mean) gives access to the mean
current, $I = e \mu /t_0$, and the second central moment (variance)
to the shot noise, $S_I = 2e^2 \mu_2 /t_0$ (for $t_0$ much larger
than the correlation time). We are also interested in the third
central moment, $\mu_3$, which gives the asymmetry of the
distribution function around its maximum (skewness). An important
difference to previous measurements of the third cumulant is that
our method can be used to extract any higher order cumulants. For
the data presented here, the accuracy of the higher cumulants is
limited by the short length of the time traces.


We first focus on the mean $\mu$ of the distribution. By measuring
$\mu$ as a function of the voltage applied on gate G1 and the bias
voltage $V$, we can construct the so-called Coulomb diamonds (see
Fig. \ref{fig3}(a)). The Coulomb diamonds describe the charge
stability of the QD, normally measured in standard transport
experiments \cite{Kouw01}. Here, we present a novel way of measuring
Coulomb blockade diamonds by time-resolved detection of the
electrons using a non-invasive charge detector. We observe clear
Coulomb blockade regions as well as regions of finite current. As we
increase the bias voltage, we see steps in the current. The first
step in Fig.~\ref{fig3}(b) (see left arrow) corresponds to the
alignment of the chemical potential of the source contact with the
ground state in the QD, and the following steps with excited states
in the QD. From the resolution of the Coulomb diamonds, we see that
the sample is stable enough such that background charge fluctuations
do not play a significant role \cite{Jung04}.


In addition to the mean, we have calculated the second and third
central moments of the electron counting statistics. These two
moments are shown in Fig.~\ref{fig3}(b) for $V_{G1}=-44$ mV as a
function of the bias voltage. The second moment (blue dotted line)
reproduces the steps seen in the current. These two moments can be
represented by their reduced quantities $\mu_2/\mu$ (known as the
Fano factor) and $\mu_3/\mu$, as shown in Fig.~\ref{fig3}(c). Both
normalized moments are almost independent of the bias voltage, and
correspond to a reduction compared to the values
$\mu_2/\mu=\mu_3/\mu=1$ expected for classical fluctuations with
Poissonian counting statistics. Super-poissonian noise
\cite{Belzig01} is not expected in our configuration.


In a QD, one expects a reduction of the moments due to the fact that
when one electron occupies the QD, a second electron cannot enter.
This leads to correlations in the current fluctuations, and to a
reduction of the noise. The reduction is maximal when the tunnel
barriers are symmetric. For an asymmetrically coupled QD, the
transport is governed by the slow barrier and the noise recovers the
value for a single tunneling barrier. The normalized moments for a
single level QD at large bias voltage can be expressed as a function
of the asymmetry of the tunneling rates, $a=(\Gamma_S -
\Gamma_D)/(\Gamma_S + \Gamma_D)$ \cite{Bagrets01}:
\begin{equation}
\frac{\mu_2}{\mu}= \frac{1}{2} \left( 1 +a^2 \right) \quad
\text{and} \quad \frac{\mu_3}{\mu} = \frac{1}{4} \left( 1 + 3a^4
\right) \text{ .} \label{muvsasym}
\end{equation}
The second central moment recovers earlier calculations of the Fano
factor in a QD \cite{Hershfield03}. We see in these equations that
both moments are reduced for a symmetrically coupled QD (i.e.
$a=0$), and tend to the Poissonian values for an asymmetrically
coupled QD (i.e. $a = \pm 1$).


\begin{figure}
\includegraphics[width=1\linewidth]{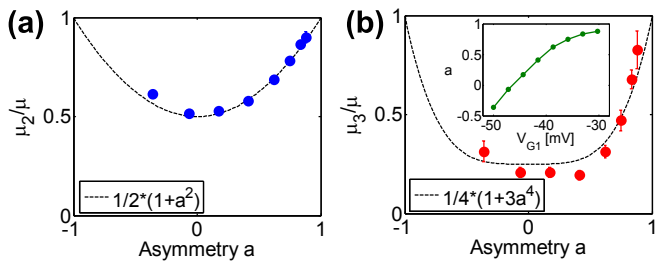}
\caption{\label{fig4} (a) Second and (b) third normalized central
moments of the fluctuations of $n$ as a function of the asymmetry of
the tunneling rates, $a=(\Gamma_S - \Gamma_D)/(\Gamma_S +
\Gamma_D)$. To increase the resolution, each point at a given
asymmetry is obtained by averaging over about $50$ points at a given
voltage $V_{G1}$ and in a window of bias voltage $1.5 < V < 3$ mV.
Error bars correspond to the standard error of this averaging
process, and are of the size of the points if not shown. The dashed
lines are the theoretical predictions given by
Eqs.~(\ref{muvsasym}). No fitting parameters have been used, since
the tunneling rates are fully determined experimentally (see
Fig.~\ref{fig1}(e) and text). Inset of (b): Variation of the
asymmetry of the tunneling rates, $a$, as a function of $V_{G1}$.}
\end{figure}

Reduction of the second moment (shot noise) due to Coulomb blockade
has already been reported in the case of asymmetrically coupled QDs
\cite{Birk02,Nauenetal}. In these experiments, reduction of the shot
noise occurs due to bias voltage dependent effective tunneling rates
\cite{Hershfield03}. Here we report the reduction of the second, as
well as the third moment for a fully controllable QD. In particular,
we are able to continuously change the tunneling rates: by changing
the gate voltage $V_{G1}$, we change the chemical potential in the
QD, and also the asymmetry of the coupling by changing the opening
of the source lead. The tunneling rates can be directly measured as
described in Fig.~\ref{fig1}(e), and the inset of Fig.~\ref{fig4}(b)
shows the variation of asymmetry with gate voltage in the region of
interest. In Fig.~\ref{fig4}(a) and \ref{fig4}(b), we show the
normalized second and third central moments as a function of the
asymmetry $a$. The experimental data follow the theoretical
predictions given by Eqs.~(\ref{muvsasym}) very well. We note in
particular that no fitting parameters have been used since the
tunneling rates are determined experimentally.


Our ability to measure the counting statistics of electron transport
relies on the high sensitivity of the QPC as a charge detector. The
counting process that we demonstrate in this paper was not possible
in previous experiments with the accuracy required for performing a
statistical analysis \cite{Fujisawa01}. Given the bandwidth of our
experimental setup, $\Delta f = 30$ kHz, the method allows to
measure currents up to $5$ fA, and we can measure currents as low as
a few electrons per second, i.e., less than 1 aA. The low-current
limitation is mainly given by the length of the time trace and the
stability of the QD, and is well below what can be measured with
conventional current meters. In addition, as we directly count
electrons one by one, this measurement is not sensitive to the noise
and drifts of the experimental setup. It is also an very sensitive
way of measuring low current noise levels. Conventional measurement
techniques are usually limited by the current noise of the
amplifiers (typically $10^{-29}$ A$^2$/Hz)
\cite{PicciottoSaminadayar,Birk02,Nauenetal}: here we demonstrate a
measurement of the noise power with a sensitivity better than
$10^{-35}$ A$^2$/Hz.

In conclusion, we have measured current fluctuations in a
semiconductor QD, using a QPC to detect single electron traveling
through the QD. We show experimentally the reduction of the second
and third moment of the distribution when the QD is symmetrically
coupled to the leads. This ability to measure current fluctuations
in a QD, as well as the very low noise level we demonstrate here,
open new possibilities towards measuring electronic entanglement in
quantum dot systems \cite{Loss02,Saraga01}.

\begin{acknowledgments}
The authors thank W. Belzig for drawing our attention to the
measurement of the full counting statistics. Financial support from
the Swiss Science Foundation (Schweizerischer Nationalfonds) via
NCCR Nanoscience and from the EU Human Potential Program financed
via the Bundesministerium f\"ur Bildung und Wissenschaft is
gratefully acknowledged.
\end{acknowledgments}

\bibliography{biblio}

\end{document}